\documentclass[twocolumn,superscriptaddress]{revtex4-2}%
\usepackage{amsmath}%
\usepackage{amsfonts}%
\usepackage{amssymb}%
\usepackage{graphicx}
\usepackage{fullpage}
\usepackage{dsfont}
\usepackage{bm}
\usepackage{multirow}
\usepackage{color}
\usepackage{float}
\usepackage{braket}

\begin{document}

\title{Revealing dark exciton signatures in polariton spectra of 2D materials}

\author{Beatriz Ferreira}
\email{beatriz.ferreira@chalmers.se}
\affiliation{Chalmers University of Technology, Department of Physics, 412 96 Gothenburg, Sweden}

\author{Hangyong Shan}
\affiliation{Institute of Physics, Carl von Ossietzky University, Oldenburg, 26129, Germany}

\author{Roberto Rosati}
\affiliation{Department of Physics, Philipps-Universit\"at Marburg, Renthof 7, D-35032 Marburg, Germany}

\author{Jamie M. Fitzgerald}
\affiliation{Department of Physics, Philipps-Universit\"at Marburg, Renthof 7, D-35032 Marburg, Germany}

\author{Lukas Lackner}
\affiliation{Institute of Physics, Carl von Ossietzky University, Oldenburg, 26129, Germany}

\author{Bo Han}
\affiliation{Institute of Physics, Carl von Ossietzky University, Oldenburg, 26129, Germany}

\author{Martin Esmann}
\affiliation{Institute of Physics, Carl von Ossietzky University, Oldenburg, 26129, Germany}

\author{Patrick Hays}
\affiliation{School for Engineering of Matter, Transport, and Energy, Arizona State University, Tempe, Arizona 85287, USA }

\author{Gilbert Liebling}
\affiliation{Institute of Applied Physics, Abbe Center of Photonics, Friedrich Schiller University Jena, 07743 Jena, Germany; Fraunhofer-Institute for Applied Optics and Precision Engineering IOF, 07743 Jena, Germany; Max-Planck-School of Photonics, 07743 Jena, Germany}

\author{Kenji Watanabe}
\affiliation{Research Center for Materials Nanoarchitectonics, National Institute for Materials Science,  1-1 Namiki, Tsukuba 305-0044, Japan}

\author{Takashi Taniguchi}
\affiliation{Research Center for Materials Nanoarchitectonics, National Institute for Materials Science,  1-1 Namiki, Tsukuba 305-0044, Japan}

\author{Falk Eilenberger}
\affiliation{Institute of Applied Physics, Abbe Center of Photonics, Friedrich Schiller University Jena, 07743 Jena, Germany; Fraunhofer-Institute for Applied Optics and Precision Engineering IOF, 07743 Jena, Germany; Max-Planck-School of Photonics, 07743 Jena, Germany}

\author{Sefaattin Tongay}
\affiliation{School for Engineering of Matter, Transport, and Energy, Arizona State University, Tempe, Arizona 85287, USA }

\author{Christian Schneider}
\affiliation{Institute of Physics, Carl von Ossietzky University, Oldenburg, 26129, Germany}

\author{Ermin Malic}
\affiliation{Department of Physics, Philipps-Universit\"at Marburg, Renthof 7, D-35032 Marburg, Germany}

\begin{abstract}
Dark excitons in transition metal dichalcogenides (TMD) have been so far neglected in the context of polariton physics due to their lack of oscillator strength. However, in tungsten-based TMDs, dark excitons are known to be the energetically lowest states and could thus provide important scattering partners for polaritons.
In this joint theory-experiment work, we investigate the impact of the full exciton energy landscape on polariton absorption and reflectance. 
By changing the cavity detuning, we vary the polariton energy relative to the unaffected dark excitons in such a way that we open or close specific phonon-driven scattering channels. We demonstrate both in theory and experiment that this controlled switching of scattering channels manifests in characteristic sharp changes in optical spectra of polaritons. These spectral features can be exploited to extract the position of dark excitons. Our work suggests new possibilities for exploiting polaritons for fingerprinting nanomaterials via their unique exciton landscape.

\end{abstract}

\maketitle

%\section{Introduction}\label{sec1}

\begin{figure}[t!]
    \centering
    \includegraphics[width=0.98\columnwidth]{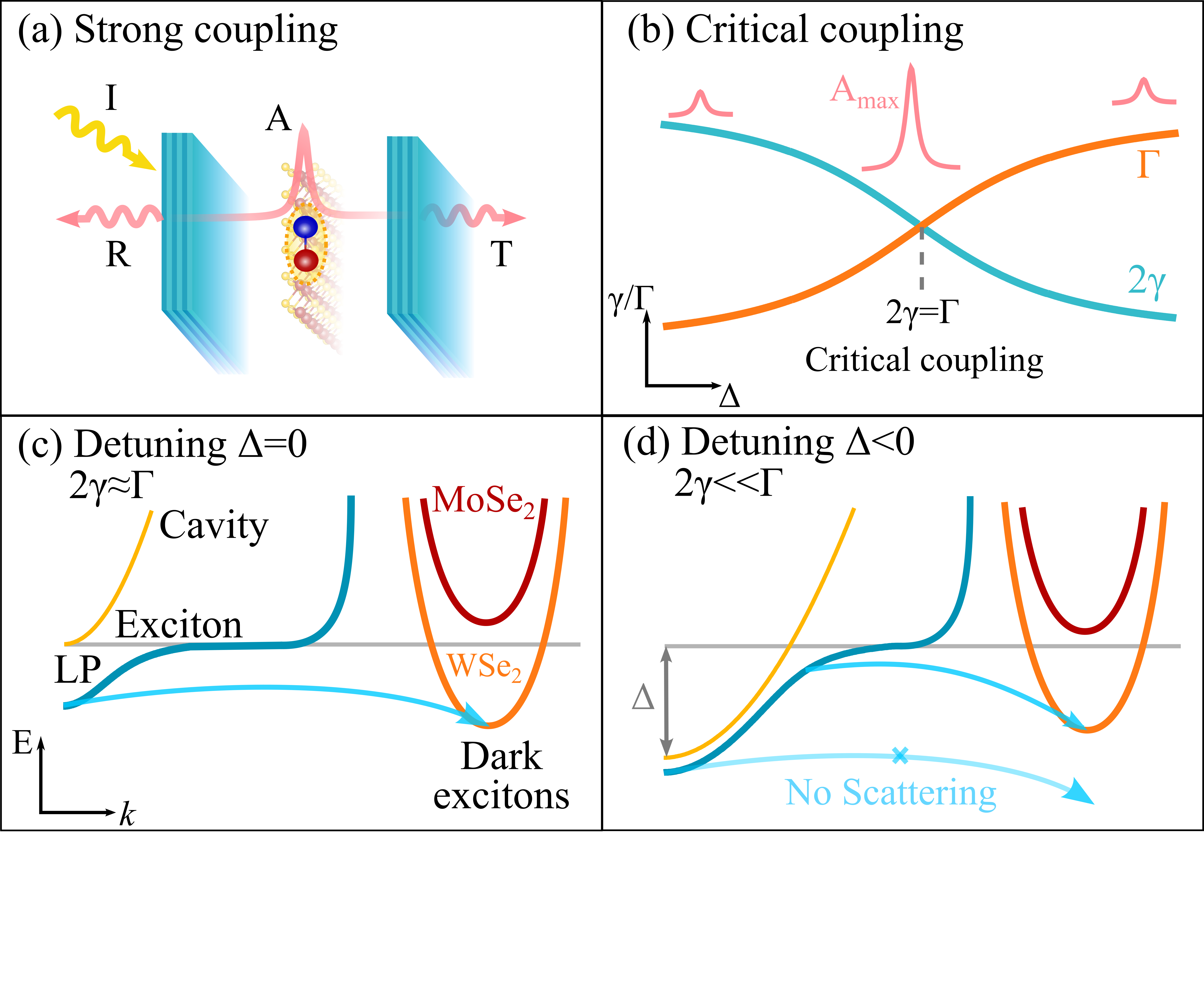}\vspace{-1cm}
    \caption{ (a) Schematic of an hBN-encapsulated TMD monolayer integrated within a Fabry-Perot microcavity. Light is shined on the left mirror of the cavity (I), which can be reflected (R), transmitted (T) or absorbed (A). (b) Critical coupling condition (reached when the total cavity decay rate equals $\Gamma$, or twice the port decay rate) results in a maximum polariton absorption. Sketch of the lower polariton branch (LP) in relation to dark exciton states in a MoSe$_2$ and a WSe$_2$ monolayer for (c)  zero detuning ($\Delta=0$) and (d) negative detuning ($\Delta<0$) between the cavity mode (yellow line) and the bright exciton state (grey line). }
    \label{fig:1}
\end{figure}

The integration of transition metal dichalcogenides (TMDs) monolayers in optical microcavities opens the door to a variety of applications in optoelectronics  \cite{liu2015,schneider2018,luo2023}. This class of atomically thin nanomaterials exhibits a large oscillator strength and exciton binding energies in the range of a few hundred of meV, hence governing optoelectronic properties even at room temperature \cite{lundt2016room,dufferwiel2015,Zhang2018,zhao23}. TMD can also form heterostructures, both vertical \cite{ciarrocchi2022,novoselov16} and lateral \cite{beret2022,rosati2023,yuan2023}. When the exciton-photon coupling strength is larger than the cavity dissipative rate and exciton non-radiative decay rate, the strong coupling regime is achieved (\ref{fig:1}(a)) \cite{savona1999,deng2010exciton}. This leads to new eigenmodes of the system, the exciton-polaritons \cite{hopfield1958theory,deng2010exciton, dufferwiel2015, zhang2021,fitzgerald2022}. Here, only bright excitons couple to light to form polaritons. In contrast, momentum-dark excitons, where Coulomb-bound electrons and holes are localized in different valleys in momentum space, possess a large centre-of-mass momentum and thus cannot be optically excited or radiatively recombine  \cite{Mueller18, Malic18, Berghauser18}.

While the importance of spin-dark excitons on optical properties of TMD-based exciton-polaritons  
has been recently considered \cite{shan2022}, the impact of momentum-dark excitons has been less studied due to their negligible oscillator strength. Nonetheless, they can still have a large influence on the scattering of polariton states \cite{ferreira2023} and, consequently, drastically change the optical response of polaritons. Here, we present a joint theory-experiment study revealing how polariton optics allows us to map the entire exciton landscape of TMD materials, including bright and dark exciton states. Based on a microscopic many-particle theory, we predict that under certain conditions, polaritons can efficiently scatter into dark exciton states, giving rise to sharp step-like changes both in polariton absorption and reflection as well as in the polariton linewidth. The prediction is confirmed in our measurements on TMD monolayers integrated into fully spectrally tunable open microcavities \cite{drawer2023}. 

Interestingly, we find that cavity detuning allows us to switch on and off phonon-driven scattering channels between the lower polariton branch and momentum-dark exciton states in tungsten-based TMDs. Depending on the detuning between the polariton and the exciton energy, realized by changing the cavity length \cite{besga2015,muller2015,gebhardt2019}, we can control the strength of the scattering rate $\Gamma$ and, consequently, the polariton absorption intensity as well as the polariton linewidth. The strongest absorption is obtained at the critical coupling condition when $\Gamma$ equals twice the cavity decay rate $\gamma$ (Fig. 1(b)) \cite{haus1984}.
For a sufficiently large detuning, the lower polariton branch is pushed below the energetically lowest dark exciton state and becomes the new ground state, closing the scattering channel between these states (Fig. 1(d)). This strategy has recently also been applied to effectively brighten a dark material and enhance its photoluminescence \cite{shan2022}.
In our work, we demonstrate the appearance of characteristic step-like increases in the detuning-resolved polariton linewidth, which can be associated with specific scattering channels into dark exciton states. This allows us to extract the energy of the involved states and map the exciton energy landscape in TMD materials.\\

\textbf{Theoretical approach:}
Using a Wannier-Hopfield method \cite{fitzgerald2022,ferreira2023,konig2023}, we model the optical response of exciton-polaritons in TMD monolayers integrated within a Fabry-Perot cavity (Fig. \ref{fig:1}(a)). The excitonic landscape is obtained by solving the Wannier equation \cite{Haug09test,berghauser14,Selig16}, including DFT input on single-particle energies \cite{Kormanyos15}. The polariton dispersion is obtained directly from the diagonalisation of the many-particle Hamiltonian in the polariton basis under the Hopfield transformation \cite{hopfield1958theory,deng2010exciton,fitzgerald2022,ferreira2023}. 
The resulting lower (LP) and the upper polariton (UP) energies can be controlled with the cavity length and with this, the detuning between the bright exciton and the cavity resonant energy, $\Delta=E_{X_0}-E_c$. Momentum-dark excitons  (KK$^\prime$, K$\Lambda$) remain unchanged as a function of detuning. 

Starting from the Heisenberg equations of motion for the coherent population of polaritons, including the external radiation field, we  obtain an Elliot-like formula for the polariton absorption for a symmetric cavity \cite{fitzgerald2022}
\begin{equation}
A^n_{\mathbf{k}}(\omega)=\frac{4\gamma_\mathbf{k}^n\Gamma^n_{\mathbf{k}}}{(\Delta^n_{\mathbf{k}})^2+(2\gamma^n_\mathbf{k}+\Gamma^n_{\mathbf{k}})^2} \quad ,
\label{abs}
\end{equation}
for each polariton branch $n$ and the in-plane momentum $\textbf{k}$. Here, we have introduced $\Delta^n_{\mathbf{k}}=\hbar\omega-E^n_\mathbf{k}$ with $E^n_\mathbf{k}$ as the polariton energy and $\hbar\omega$ as the energy of the incoming light. The polariton-phonon scattering rates $\Gamma^n_{\mathbf{k}}$ are obtained within the second-order Born-Markov approximation \cite{Lengers21,Ferreira2022}. These rates are dominated by the scattering into exciton states outside the light-cone, crucially including also the momentum-dark valleys, while scattering within the light-cone is small due to the reduced density of state \cite{Ferreira2022,ferreira2023}. In this work, we focus on spin-conserving phonon-induced processes.  The expression for the polariton absorption corresponds to the exciton Elliot formula, but with renormalized exciton-phonon matrix elements \cite{thranhardt2000,Brem18} now including the excitonic Hopfield coefficients \cite{Ferreira2022,ferreira2023}. The total radiative decay rate of polaritons $2\gamma^n_{\mathbf{k}}$, i.e. the sum of the photon leakage through both ports, is given by the total cavity decay rate scaled by the photonic Hopfield coefficient for state $n$ with momentum $\mathbf{k}$ \cite{gardiner1985,fitzgerald2022}. Hence, both main contributions for the absorption scale with the two Hopfield coefficients. The maximum value of absorption, $A=0.5$, is obtained at the critical coupling condition \cite{adler1960,haus1984} of $2\gamma^n_\mathbf{k}=\Gamma^n_\mathbf{k}$, i.e. when the leakage out of both ports of the cavity is equal to the exciton dissipation rate  \cite{piper2014}, cf. Fig. \ref{fig:1}(b). The absorption equation (\ref{abs}) has a Lorentzian-like shape with the full width at half maximum  $2(2\gamma^n_{\mathbf{k}}+\Gamma^n_{\mathbf{k}})$ corresponding to the polariton linewidth. This quantity will be used to directly compare the theoretical prediction with experimental results later in the text.\\

\begin{figure}[t!]
    \centering
    \includegraphics[width=\columnwidth]{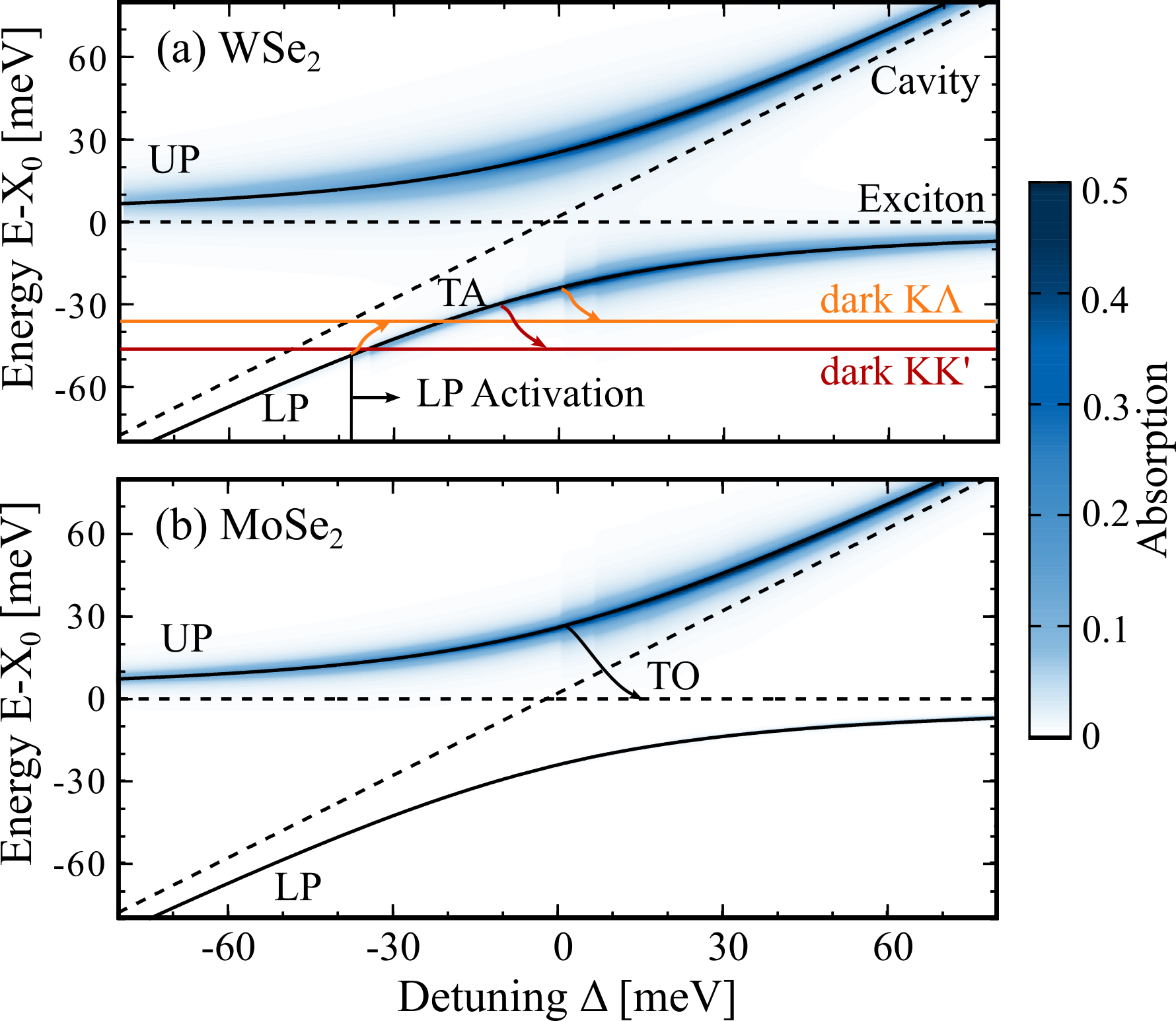}
    \caption{Polariton absorption as a function of detuning and energy (in relation to the bright exciton energy X$_0$) at zero in-plane momentum  for the upper (UP) and the lower polariton (LP) for (a) WSe$_2$ and (b) MoSe$_2$ monolayers at 77 K.}
    \label{fig:2}
\end{figure}

\textbf{Polariton Absorption:}
We start by studying the detuning-resolved polariton absorption spectra at 77K, calculated by numerically evaluating Eq. (\ref{abs}), for hBN-encapsulated WSe$_2$ and MoSe$_2$. 
We consider the case of a symmetric cavity and the TMD integrated into the center of the cavity.  The cavity parameters have been chosen to match the experimental cavity decay rate in the full photonic limit. Furthermore, we consider a realistic Rabi-splitting of 50 meV \cite{schneider2018}, and we focus on light emitted at normal incidence, corresponding to an in-plane momentum $k=0$. 
Interestingly, we find a drastic difference in the polariton absorption of tungsten- and molybdenum-based TMDs, particularly considering the lower polariton branch, cf. Fig. \ref{fig:2}. 
As the cavity decay rates for WSe$_2$ and MoSe$_2$ are similar, the predicted difference in the absorption clearly stems from the polariton-phonon scattering rate $\Gamma$. 

In MoSe$_2$, we find only a negligibly small absorption of the lower polariton branch, cf. Fig. \ref{fig:2}(b).  In stark contrast, WSe$_2$ exhibits negligibly LP absorption only for detuning values below $\Delta=-38$ meV. Above this value, the LP absorption first slowly increases and shows then a relatively sharp increase at $\Delta$=-10 and 0 meV. These values of detuning correspond to polaritonic energies of approximately 16 and 12 meV above the dark exciton state, i.e. large enough to scatter into KK' and K$\Lambda$ excitons via emission of intervalley phonons, respectively (red and orange lines in Fig. 2(a)).  The last increase is larger given the stronger scattering toward K$\Lambda$ excitons, induced by its three-fold degeneracy. This leads to an increase in both the polariton linewidth and polariton absorption (darker blue area) as the absorption approaches the critical coupling condition (Fig. \ref{fig:1}(b)).
Note that the slow increase starting at approximately -40 meV reflects the scattering into excitons via absorption of intervalley phonons. 

The upper polariton spectra are more similar for tungsten and molybdenum monolayers. We find a significant polariton absorption for all detuning values for both materials. The reason is that for the upper polariton branch, the acoustic intravalley scattering is always allowed, contrary to the case of the LP branch \cite{Ferreira2022}. The additional intervalley scattering channels into dark exciton states in WSe$_2$ give rise to only a slight increase in the absorption magnitude. Since exciton-phonon scattering is more efficient in MoSe$_2$ than in WSe$_2$ \cite{Jin14}, we find a broader linewidth and height when the intravalley scattering induced by optical emission opens up in MoSe$_2$ above $\Delta\approx$ 10 meV (Fig. \ref{fig:2}(b)). 

\begin{figure}[t!]
    \centering
    \includegraphics[width=\columnwidth]{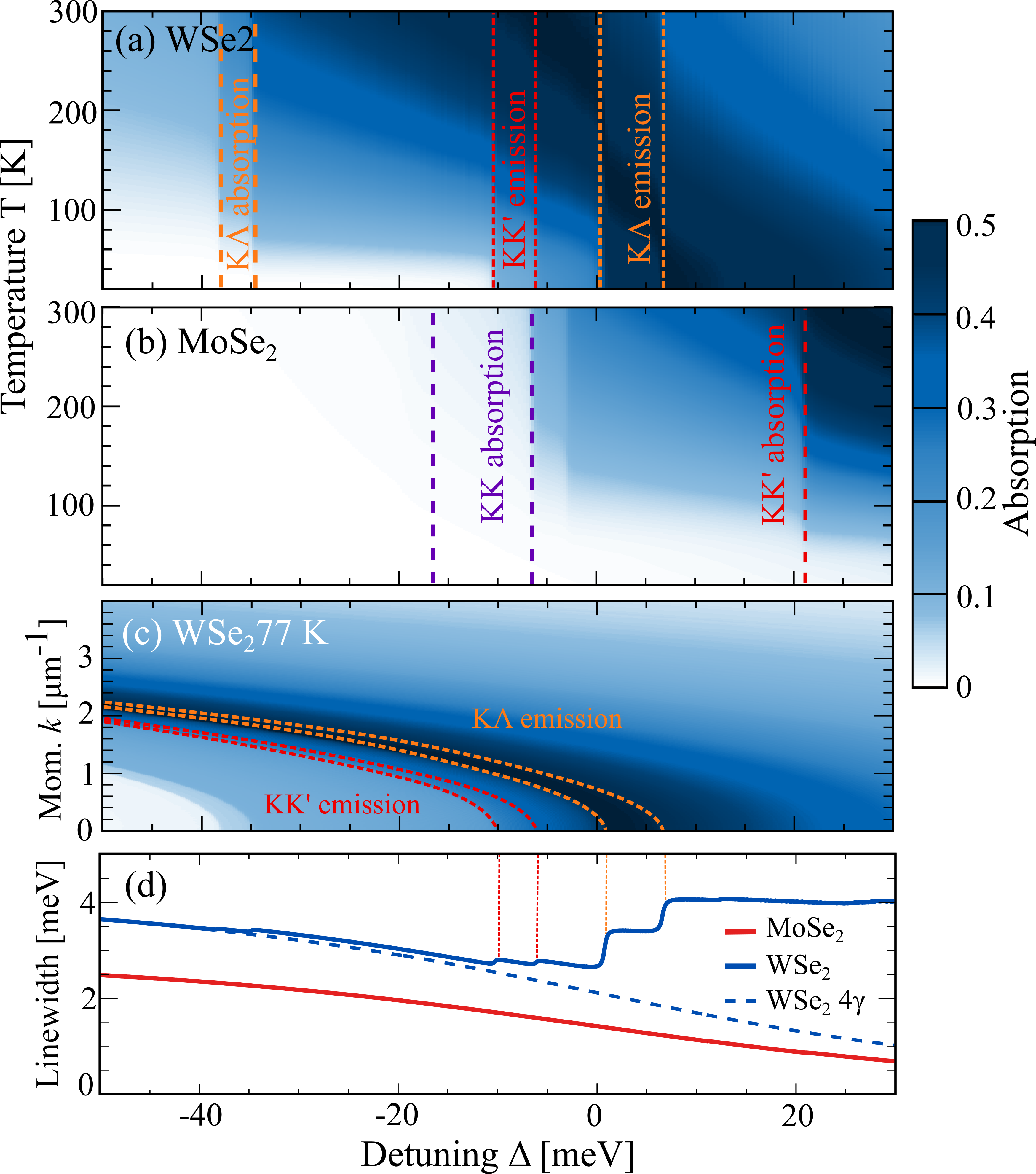}
    \caption{Lower polariton absorption as a function of detuning and temperature for  (a)  WSe$_2$ and (b)  MoSe$_2$ monolayers in a cavity. (c) Polariton absorption of WSe$_2$ at the fixed temperature of 77K as a function of detuning and momentum. The absorption increases as more phonon-driven scattering channels into the dark exciton states K$\Lambda$ and KK$^\prime$ open up, as denoted by the vertical orange and red lines in (a). (d) Detuning-resolved lower polariton linewidth at 77 K.}
    \label{fig:3}
\end{figure}

Temperature plays a crucial role in polariton absorption as it dictates the efficiency of phonon-driven scattering processes \cite{Brem19,ferreira2023}. With this in mind, we investigate the polariton absorption as a function of detuning and temperature, both for WSe$_2$ and  MoSe$_2$ monolayers in a cavity, cf. Figs. \ref{fig:3}(a) and (b). We find that, interestingly, both materials show regions with negligible polariton absorption (indicated by white areas). In WSe$_2$, we predict three regions with relatively sharp increases in the absorption magnitude that are triggered by the opening of specific scattering channels. We find increased polariton absorption around $\Delta$=-10 and 0 meV, which is driven by the scattering into the dark KK$^\prime$ and K$\Lambda$ excitons, respectively, via emission of LA and TA phonons.
The dashed vertical lines indicate the minimum LP energy required for scattering with LA and TA phonons into the corresponding dark exciton state. Note that these features are relatively weakly temperature-dependent, and they are still visible even at room temperature. We also find a third relatively sharp increase in the polariton absorption for a detuning of approximately -40 meV, now with a stronger temperature dependence. The enhanced absorption starts to become visible only for temperatures larger than approximately 100K. This occurs as the scattering into the K$\Lambda$ states is only allowed via the \textit{absorption} of phonons, which is activated at sufficiently large temperatures. Note that phonon absorption into the KK' states is possible even for smaller detuning values ($\Delta=-55$  for LA absorption), as this valley lies below the K$\Lambda$ state. Thus, there is no additional increase visible in the presented detuning range. 

Overall, the absorption in MoSe$_2$ is weaker than in WSe$_2$ due to the reduced effects of dark excitons supported by the different band alignment, which forbids the scattering via emission of phonons. At high temperatures, a few weak step-like increases are present due to the thermally-activated absorption of intravalley optical modes (purple) or intervalley phonons toward the dark KK' valley (red). In MoSe$_2$, the scattering into KK' is possible only at high detunings because the energy of KK' states is 9 meV above the bright exciton, in contrast to WSe$_2$ where it lies 46 meV below the bright exciton (cf. Fig. \ref{fig:1}(c)).

So far, we have investigated the polariton absorption at zero momentum. Figure \ref{fig:3}(c) shows the LP absorption in WSe$_2$ at a fixed temperature of 77K as a function of detuning and momentum. For negative cavity detuning values and below $k=2$ $\mu$m$^{-1}$, there is only a negligibly small absorption. The reason is that there is a certain polariton access energy required to scatter into the dark exciton states (Fig. \ref{fig:1}(d)). Furthermore, the steep lower polariton dispersion allows us to show the opening of these scattering channels as a function of momentum. The smaller the negative detuning, the lower is also the required momentum for this scattering to occur with critical detuning values for different channels appearing at $k=0$ $\mu$m$^{-1}$, cf. Fig. \ref{fig:3}(c).

Finally, we show that not only the absorption magnitude but also the spectral linewidth of the lower polariton [$2(2\gamma+\Gamma)$] is strongly sensitive to the opening of the detuning-driven polariton-phonon scattering channels, cf. Fig. \ref{fig:3}(d).
For MoSe$_2$, due to the limited polariton-phonon scattering channels, the LP linewidth is determined by the cavity decay rate $4\gamma$, i.e. on the photonic Hopfield coefficient of the lower polariton, which decreases with increasing detuning. As a direct consequence, we find a reduced LP linewidth as the detuning increases (red line in Fig. \ref{fig:3}(d)). We find the same behaviour also for WSe$_2$ until the polariton-phonon scattering channels start to open up at specific detuning values, resulting in a series of sharp increases of the LP linewidth.  \\

\textbf{Polariton linewidth:}
Now, we complement the theoretical predictions with cryogenic reflectance measurements of WSe$_2$ and MoSe$_2$ monolayers in a cavity. We generalize Eq. (\ref{abs}) to study the reflected signal given the asymmetric cavity-decay rates in the two mirrors of our cavity.  We utilize the full spectral tunability of our open cavity to map the coupling conditions of the two considered samples based on integrated MoSe$_2$ and WSe$_2$ monolayers. Both experiments yield a clear anti-crossing behaviour of the upper- and lower polariton branches in the reflection contrast study, featuring a Rabi-splitting of 20 meV for WSe$_2$ and a slightly reduced Rabi-splitting of 16 meV for MoSe$_2$   
To assess the role of dark excitons in our experiment, we extract the linewidth of the lower polariton resonance by fitting it with a Fano-type of lineshape and plot the result as a function of detuning. 

\begin{figure}[t!]
    \centering
    \includegraphics[width=\columnwidth]{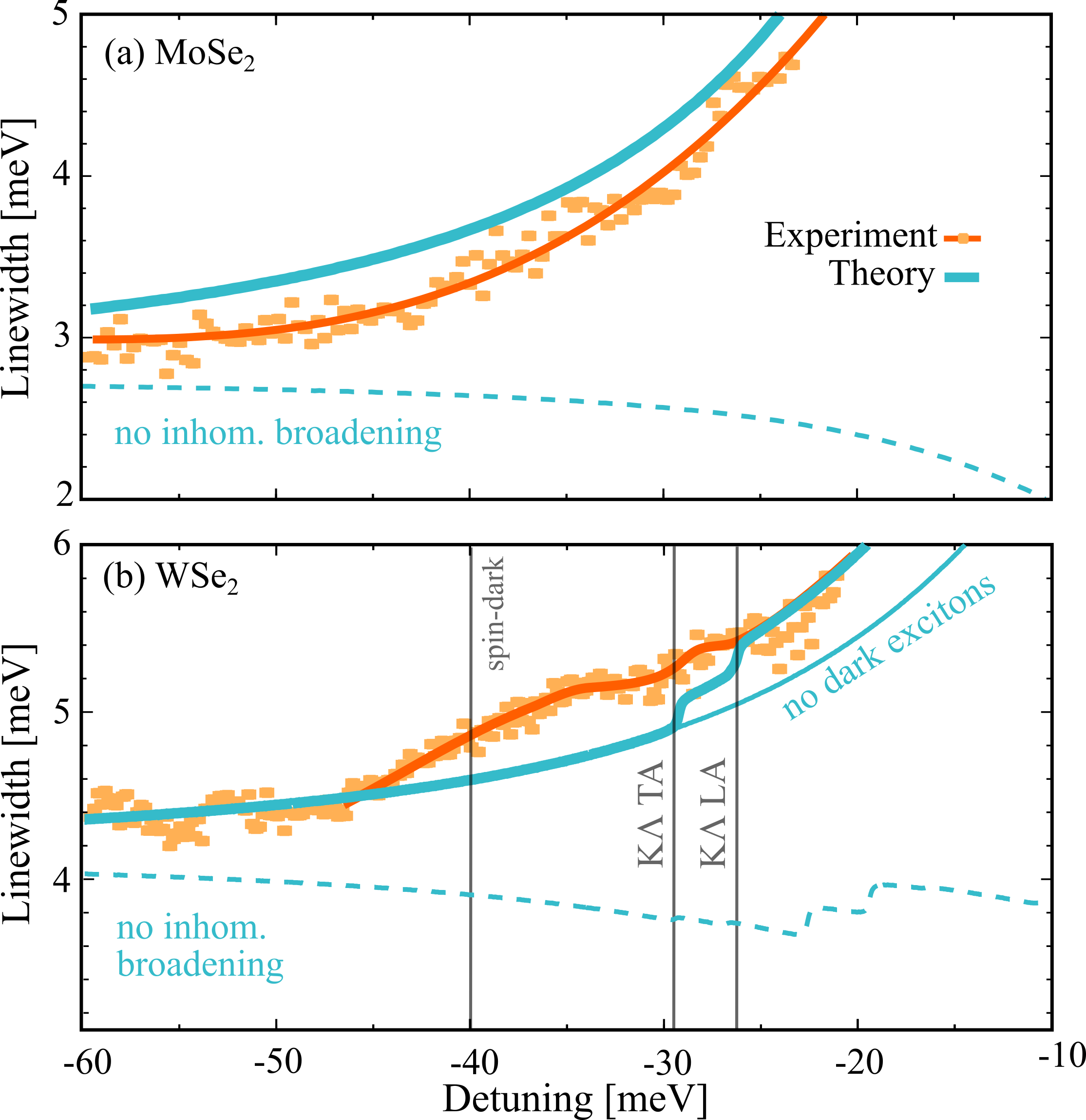}
    \caption{Direct experiment-theory comparison. Theoretically predicted (blue line) and experimentally measured (orange dots) lower polariton linewidth obtained from reflection measurements at 20 K for a (a)  MoSe$_2$  and (b)  WSe$_2$ monolayer in a cavity. The thin blue line shows the calculated linewidth without scattering into dark exciton states, while the dashed line represents the intrinsic linewidth without inhomogeneous broadening. The vertical black lines denote the energy of spin-dark excitons in experiments as well as the detuning values at which the scattering channels with TA and LA phonons are opened.}
    \label{fig:4}
\end{figure}

Figure \ref{fig:4} shows a direct comparison between the theoretically predicted and experimentally measured lower polariton linewidth as a function of detuning at a fixed temperature of 3.5 K. While for MoSe$_2$, we observe both in theory and experiment a smooth increase of the LP linewidth as a function of detuning (Fig. \ref{fig:4}(a)), we find a more complex behaviour for WSe$_2$ (Fig. \ref{fig:4}(b)). We observe in the experiment a series of step-like increases in the polariton linewidth at the specific detuning values of approximately -40, 29 and 26 meV  (orange line in Fig. \ref{fig:4}(b)). Our theoretical predictions can reproduce well the latter two  (blue line) and can clearly trace back their microscopic origin to the opening of the scattering channel toward the dark K$\Lambda$ exciton via the emission of TA and LA phonons, respectively (cf. also Fig. \ref{fig:2}(a)-(b)). When neglecting the scattering to dark states, these characteristic sharp increases vanish (thin blue line in Fig. \ref{fig:4}(b)). Note that to match the experiments better, we have slightly red-shifted the microscopically obtained energy of K$\Lambda$ excitons (by 6meV). The exact position of these dark states is still under debate in literature.  In our microscopic model, we include only spin-conserving phonon-induced scattering processes. It is likely that the interaction with the spin-dark KK excitons is responsible for the first increase in the linewidth,  whose broad shape is roughly centred around the energy of these states (detuning of -40 meV). 

Note that the direct comparison between theory and experiment suggests that inhomogeneous broadening plays an important role, as indicated by the overall increase of the polariton linewidth with detuning also in the $\Delta \ll 0$ regime (solid lines in Fig. \ref{fig:4}). In contrast, for a purely homogeneous linewidth, we predict a decrease with detuning, the dashed blue line in Fig. \ref{fig:4}, reflecting the behaviour of the photonic Hopfield coefficient. This suggests the existence of macroscopic/micrometric inhomogeneities, e.g. due to strain in the sample, whereas smaller-scale disorder could lead to motional narrowing \cite{Wurdack21}.  Hence, we have introduced an inhomogeneous broadening enhancing the exciton scattering in the rate $\Gamma$  to match the general experimental trend with detuning. This has no impact on the main message of our work, namely the appearance of the characteristic series of step-like increases in the polaritonic linewidth at specific detuning values.  Their observation in Fig. \ref{fig:4} has allowed us to experimentally determine the relative energy of bright and dark excitons in TMD materials via polaritonic spectra. \\

\textbf{Conclusion:}
We have presented a microscopic study combining material-specific many-particle theory with cryogenic spectroscopic measurements to investigate polariton absorption and reflectance in hBN-encapsulated WSe$_2$ and MoSe$_2$ monolayers integrated into a tunable Fabry–Perot cavity.
Interestingly, we find both in theory and experiment the appearance of a series of characteristic step-like increases at specific detuning values both in polariton absorption and polariton linewidth in WSe$_2$ monolayers. This behaviour is a clear signature of the opening of phonon-driven scattering channels between the lower polariton branch and the dark exciton states in WSe$_2$. These results show the potential of polaritons to map the full exciton landscape, including bright and dark states in atomically thin semiconductors.\\

\textbf{Acknowledgements:}
We acknowledge funding from the Deutsche Forschungsgemeinschaft (DFG) via SFB 1083 and the regular project 524612380, as well as from the Knut and Alice Wallenberg Foundation via the Grant KAW 2019.0140. The computations were enabled by resources provided by the Swedish National Infrastructure for Computing (SNIC).  
We also acknowledge financial support within the projects SCHN1376 11.1 and SCHN1376 14.1, and within the program for major equipment (INST 184/220-1 FUGG) funded by the German Research Foundation (DFG) is acknowledged.  Support by the Niedersächsisches Ministerium für Wissenschaft und Kultur (MWK) within the project "DyNano" is gratefully acknowledged.
B.H. gratefully acknowledges support from the Alexander von Humboldt Foundation.  
K.W. and T.T. acknowledge support from the JSPS KAKENHI (Grant Numbers 21H05233 and 23H02052) and World Premier International Research Center Initiative (WPI), MEXT, Japan.
S.T acknowledges primary support from DOE-SC0020653 (materials synthesis), Applied Materials Inc., NSF CMMI 1825594 (NMR and TEM studies), NSF DMR-1955889 (magnetic measurements), NSF CMMI-1933214, NSF 1904716, NSF 1935994, NSF ECCS 2052527, DMR 2111812, and CMMI 2129412.

\bibliography{Mainarxiv}

\end{document}